\documentclass{article}
\pdfoutput=1

\usepackage{PRIMEarxiv}

\usepackage[utf8]{inputenc} 
\usepackage[T1]{fontenc}    
\usepackage[hidelinks]{hyperref}       
\usepackage{url}            
\usepackage{booktabs}       
\usepackage{amsfonts}       
\usepackage{nicefrac}       
\usepackage{microtype}      
\usepackage{lipsum}
\usepackage{fancyhdr}       
\usepackage{graphicx}       

\usepackage{natbib}

\usepackage{multirow}
\usepackage{floatrow}
\newfloatcommand{capbtabbox}{table}[][\FBwidth]
\usepackage{blindtext}
\usepackage{balance}
\usepackage{caption}
\usepackage{adjustbox}

\usepackage[dvipsnames]{xcolor}
\usepackage{pgfplots} 
\pgfplotsset{compat=1.17}

\usepackage{subcaption}
\usepackage{orcidlink}

\usepackage{color,soul} 
\usepackage[toc,page]{appendix} 
\usepackage{floatrow}

\definecolor{myblue}{RGB}{68, 119, 170}
\definecolor{mycyan}{RGB}{102, 204, 238}
\definecolor{mygreen}{RGB}{34, 136, 51}
\definecolor{myyellow}{RGB}{204, 187, 68}
\definecolor{myred}{RGB}{238, 102, 119}
\definecolor{mypurple}{RGB}{170, 51, 119}
\definecolor{mygrey}{RGB}{187, 187, 187}

\usepackage{algpseudocode}
\usepackage{algorithm}

\usepackage{listings}
\usepackage{authblk}

\title{Skin Cancer Machine Learning Model Tone Bias
}

\author[1]{James Pope\orcidlink{0000-0003-2656-363X}}
\author[1]{Md Hassanuzzaman\orcidlink{0000-0002-4751-3773}}
\author[4]{William Chapman}
\author[4]{Huw Day\orcidlink{0000-0002-0744-1576}}
\author[2]{Mingmar Sherpa}
\author[1]{Omar Emara}
\author[3]{Nirmala Adhikari}
\author[1]{Ayush Joshi\orcidlink{0000-0001-9717-0724}}

\affil[1]{University of Bristol, Intelligent Systems Laboratory, Bristol, United Kingdom}
\affil[2]{Massachusetts Institute of Technology, Department of Biology, Cambridge, United States}
\affil[3]{University of Alabama at Birmingham, Department of Physics, Birmingham, Alabama, United States}
\affil[4]{Jean Golding Institute, University of Bristol, United Kingdom}

\begin{document}
\maketitle

\begin{abstract}

\textbf{Background:} Many open-source skin cancer image datasets are the result of clinical trials conducted in countries with lighter skin tones. Due to this tone imbalance, machine learning models derived from these datasets can perform well at detecting skin cancer for lighter skin tones. Though there is less prevalence of skin cancer with darker tones, any tone bias in these models would introduce fairness concerns and reduce public trust in the artificial intelligence health field.

\textbf{Methods:} We examine a subset of images from the International Skin Imaging Collaboration (ISIC) archive that provide tone information.  The subset has more light (83.3\%) than dark (16.7\%) images and more benign (74.7\%) than malignant (25.3\%) images.  These imbalances could explain a model's tone bias.  To address this, we train models using the imbalanced dataset (3,623 images) and a sampled balanced dataset ($\sim$500 images) to compare against.  The datasets are used to train a deep convolutional neural network model to classify the images as malignant or benign.  We then evaluate the models' disparate impact, based on selection rate, relative to dark or light skin tone.

\textbf{Results:} Using the imbalanced dataset, we found that the model is significantly better at detecting malignant images in lighter tone (selection rate 27.5\%) images versus darker tones (selection rate 15.9\%).  This results in a disparate impact of 0.577.  Using the balanced dataset, we found that the model is also significantly better at detecting malignant images in lighter (selection rate 50.0\%) versus darker tones (selection rate 34.2\%).  This results in a disparate impact of 0.684.  Using the imbalanced or balanced dataset to train the model still results in a disparate impact well below the standard threshold of 0.80 which suggests the model is biased with respect to skin tone.

\textbf{Conclusion:} The results show that typical skin cancer machine learning models can be tone biased.  These results provide evidence that diagnosis or tone imbalance is not the cause of the bias.  These results provide evidence the models are learning tone related features. Other techniques will be necessary to identify and address the bias in these models, an area of future investigation.

\end{abstract}


\section{Introduction}
\label{sec:introduction}

In the past decade numerous skin image datasets have been publicly released \cite{CASSIDY2022} and used for developing models to predict skin diseases, including skin cancer \citep{JAIN2021, LIU2020}. However, most of the datasets are sampled from lighter skin tone populations \citep{LANCET2022}.  This raises concerns about the developed models being biased towards lighter skin tones.






Recent work on machine learning bias proposes methods to detect and mitigate model bias \citep{AIBIAS2024}.  Machine learning models utilise relationships between the input features and the target feature to make predictions.  There are concerns when the input features include or are strongly correlated with \textit{protected features} (\cite{BAROCAS2023}), also known as \textit{protected characteristics}.  These protected features include race, age, gender, nationality, and religion.  In certain applications it is acceptable to use these features to improve model predictions, such as healthcare applications \cite{BONHAM2016}.

However, in many applications using protected features to make predictions is highly undesirable and often illegal (such as finance).  Existing research regarding human diagnosis shows that healthcare workers familiar with lighter skin tones are poor at diagnosing skin cancer in darker skin tones \cite{MINDGAP2020, BARATA2023}.  Though tone is not usually included as a protected feature, it can be used as a proxy.  Even if this is acceptable, detecting and understanding the source of tone bias motivates this study.

Deep convolutional neural network models (CNNs) have been shown to perform well for skin cancer diagnosis \cite{JAIN2024}.  Given a dermoscopic image of a lesion, the model is trained to predict malignant or benign. Though the models often only use features derived from the image, there is concern the models are learning tone, or tone correlated, features.  Determining what deep learning models have learned is difficult and remains an active area of research.  Additionally, less than 2.1\% of the datasets provide skin tone annotations \cite{LANCET2022} (usually based on the Fitzpatrick skin type \cite{GUPTA2019}).



In this study, we use the publicly available International Skin Imaging Collaboration (ISIC) datasets \cite{ISIC_Archive}. We consider the subset of dermoscopic images that provide Fitzpatrick skin type.  This derived dataset has three times more light tone images than dark.  We develop and train deep CNN models on this tone imbalanced dataset to predict the diagnosis and find that it does have a tone bias.  Believing the imbalance is the source of the bias, we train models on tone balanced datasets and find that the bias remains. The results provide evidence that the model is learning tone related features.  The imbalanced dataset was limited to only dermoscopic images with skin tone information (3,623 of $\sim$81,000), greatly reducing the number of images considered. Future studies will consider using a tone classifier to utilise all dermoscopic images and explainable AI techniques to determine what the model learned regarding tone.  Future investigation will also include exploring these approaches to potentially detect skin abnormalities related to the presence of heavy metal contamination in the environment which present risk for neurotoxicity, reducing life expectancy. \cite{NIRM2024}.


\section{Material and Methods}
\label{sec:methods}
In this section we first provide details about the study approach and datasets used. We consider model bias and evaluation measures.  We then present details about the deep CNN model used in the study followed by the experimental setup.



\subsection{Study design and population}

The study here is retrospective design that harness the work of deep convolutional neural network model to classify the images as malignant or benign. We used the machine learning approach where we trained the model using the images from publicly available International Skin Imaging Collaboration (ISIC) dataset. The model is trained to perform the specific classification of the images as malignant or benign. The images data comprises from patients globally, ensuring a wide representation of demographics across the world. The diversity in skin images allow for generalisation of our model across different population. Patient information such as name, age, sex and ethnicity are anonymised in the ISIC dataset, where we ensure the research ethical standards.  The images with skin tone came from the following institutes

\begin{itemize}
    \item 1517 Hospital Italiano de Buenos Aires
    \item 1459 Sydney Melanoma Diagnostic Center at Royal Prince Alfred Hospital, Pascale Guitera
    \item 709 Memorial Sloan Kettering Cancer Center
\end{itemize}

Though this does not provide global representation, it does provide a more diverse patient population.  Future work will be to use 81,000 images that will include an even larger patient population.

\subsection{Dataset Ethics}

Our study uses images publicly available under Creative Commons licenses from the International Skin Imaging Collaboration website \cite{ISIC_Archive}. To the best of our knowledge, the images used in this study do not have personally identifiable information and are in compliance with current US Health Insurance Portability and Accountability Act (HIPPA) laws and EU General Data Protection Regulations.

\subsection{Dataset}

\begin{figure}
    \centering
    \includegraphics[width=0.8\textwidth]{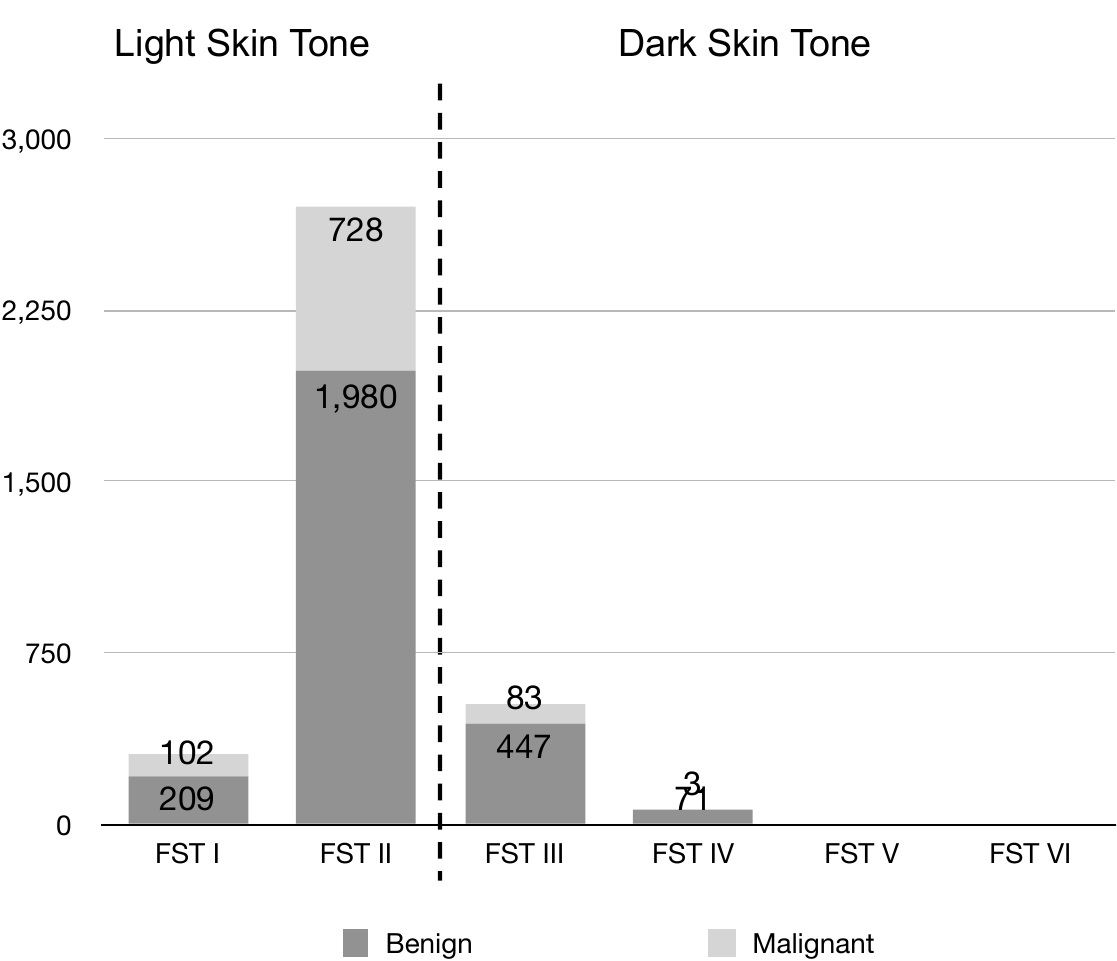}
        \caption{Dataset Fitzpatrick Skin Types to Tone Mapping}
        \label{fig:dataset_skin_types}
\end{figure}

The International Skin Imaging Collaboration has over 20 datasets from around the world and included $\sim$81,155 dermoscopic images with the diagnosis of \{\textit{malignant},\textit{benign}\}.  Of this, there are 3,623 images that also have the Fitzpatrick Skin Type (FST).  Figure \ref{fig:dataset_skin_types} shows the number of instances for each skin type and delineated by diagnosis.  There are clearly far more FST II than any other skin type.  Notably, there are not any images in the ISIC archives that have FST V or FST VI.  Based on the available images and annotated skin types, we decide to map FST I and FST II to \textit{light skin tones} and FST III and FST IV to \textit{dark skin tones}.  This decision is somewhat arbitrary.  Future will explore skin tone classification approaches using individual typology angle (ITA) \cite{CHARDON1991}.

\begin{figure}
    \centering
    \includegraphics[width=0.4\textwidth]{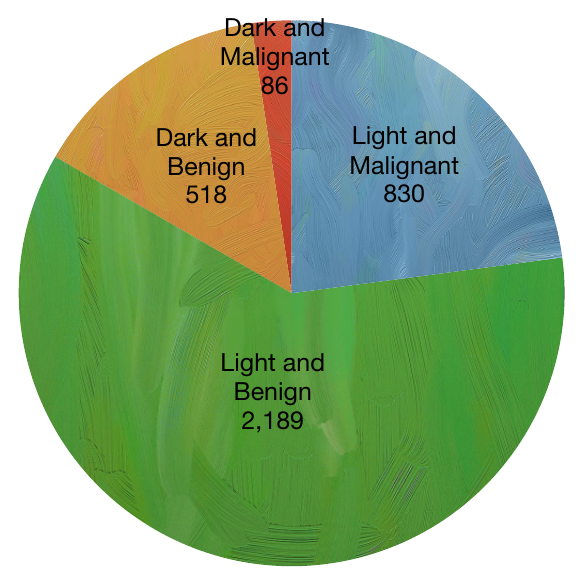}
    \caption{Imbalanced Dataset}
    \label{fig:imbalanced}
\end{figure}

Once the image skin tones have been mapped, Figure \ref{fig:imbalanced} shows the relative number for tone and diagnosis.  There is clearly a class imbalance regarding the diagnosis with 74.7\% (2707/3623) benign and  25.3\% (916/3623) malignant.  Furthermore, there is also an imbalance between tone with 83.3\% light (3019/3623) and 16.7\% light (604/3623).


\begin{figure}
    \centering
    \includegraphics[width=0.3\textwidth]{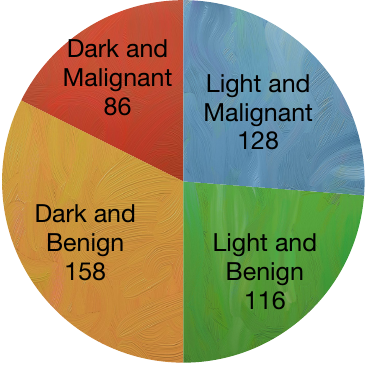}
    \caption{Balanced Dataset (Sampled)}
    \label{fig:balanced}
\end{figure}


Both imbalances are a problem.  The imbalance with diagnosis, the target feature being predicted, can cause a classifier to concentrate on this majority class and fail to learn how to classify the minority class.  This is not a bias issue but still undesirable.  The skin tone imbalance can similarly result in a classifier that is better at classifying malignant lesions for lighter tones and this potential bias is the focus of our work.  There are a number of solutions to address these imbalances, such as under-sampling the majority class, over-sampling the minority class (similar to bootstrapping), and generative/augmentation approaches.  We decide to under-sample benign images so that there are the same number of benign and malignant images.  This results in much fewer images.  The resulting dataset will have 2 $\times$ the number of malignant images, much fewer than the \textit{imbalanced dataset}.  We further under-sample light images to match dark tone images.  This slightly changes thee ratio for diagnosis.  Figure \ref{fig:balanced} shows the result after under-sampling by diagnosis and skin tone.  Clearly the skin tone is perfectly balanced and diagnosis is much more balanced than before.  However, there are far fewer images in the balanced dataset versus the imbalanced dataset.  We will use both these datasets to train models and evaluate bias.



\subsection{Evaluation Metrics}

In this section we first define terms used in the \textit{confusion matrix}, followed by common evalaution metrics, and finish with a discussion of which metrics we chose.

\subsubsection{Confusion Matrix Definition}

For a given image, a binary skin cancer classification model will predict either \textit{benign} or \textit{malignant} and can be compared against the true diagnosis for the image.  For binary classification, the more general terms are \textit{positive} and \textit{negative}.  We define \textit{positive} to indicate \textit{malignant} and define \textit{negative} to indicate \textit{benign}.  With these definitions, we further define the following terms.

\begin{itemize}
    \item \textit{true positive} (TP) the classifier predicts \textit{malignant} and the true diagnosis is \textit{malignant},
    \item \textit{true negative} (TN) the classifier predicts \textit{benign} and the true diagnosis is \textit{benign},
    \item \textit{false positive} (FP) the classifier predicts \textit{malignant} and the true diagnosis is \textit{benign},
    \item \textit{false negative} (FN) the classifier predicts \textit{benign} and the true diagnosis is \textit{malignant}.
\end{itemize}

Given a large number of images to predict, the first step in evaluating the model is to compute number of true positives, true negatives, false positives, false negatives.  These four terms are usually presented in a \textit{confusion matrix} with the true predictions on the diagonal and the false predictions off-diagonal.  This presentation is useful and we use it in our \ref{sec:results} section.

\subsubsection{Model Evaluation Metric}

The \textit{confusion matrix} is used to compute several metrics that can be used to evaluate the performance of the model. Given the confusion matrix, the accuracy is defined as follows.

\begin{equation}
\label{equ:accuracy}
\textit{accuracy} = \frac{TP + TN}{TP + TN + FP + FN}
\end{equation}

The accuracy includes all terms and measures the number of times the model is correct versus the number of predictions. For a balanced dataset, where the number of \textit{malignant} images is roughly equal to the \textit{benign} images, accuracy is a useful metric to evaluate the performance of a classifier.  However, when there are significantly more \textit{benign} than \textit{malignant} images in the datasets, known as a \textit{class imbalance}, accuracy is not best choice.  This is because a trivial \textit{majority class classifier}, that just predicts benign, can achieve a high  accuracy.  Other scores consider different combinations of TP, TN, FP, and FN (such as precision, recall, and f1) and are also often used (for both balanced and imbalanced datasets).  

For our purposes, we are concerned with the disparate impact (also derived from TP, TN, FP, and FN). However, regarding accuracy, the model needs to perform better than the majority class classifier. For the imbalanced dataset, the \textit{majority class classifier} would achieve an accuracy of 74\%.  Any model that cannot achieve a higher accuracy would indeed be considered poor.  A model that is learning to differentiate between \textit{malignant} and \textit{benign} should result in an accuracy higher than the majority class percentage.

\subsubsection{Model Training}

The number of epochs (i.e. how long to train the model) can greatly affect the accuracy of a model.  During each epoch the training loss is generally expected to decrease.  A model is considered \textit{trained} when the training loss ceases to decrease.  The model is said to have learned all it can from the dataset.   The training stops using some criteria.  We decide to stop when the training loss generally stops decreasing.  This stopping criteria is generally termed \textit{early stopping} (this can also combined with differences between the training and validation losses).  Importantly, this approach is often used and would be typical of a deployed model.  We provide a detailed bias analysis using the trained models.

\subsubsection{Evaluating Model Bias}
\label{sec:bias_evaluation}

\begin{table}
    \centering
    \begin{tabular}{|c|c|c|}
       \hline
       Criterion  & Statements  & Condition \\
       
       \hline
         &  &  \\
       
       Independence  & 
       $\hat{Y} \perp A$ & 
       $\mathbb{P} \{\hat{Y} = 1 \mid A = a\} = \mathbb{P} \{\hat{Y} = 1 \mid A = b\}$  \\

         &  &  \\
       
       Separation    & 
       $\hat{Y} \perp A \mid Y$ & 
       $\mathbb{P} \{\hat{Y} = 1 \mid Y = 1, A = a\} = \mathbb{P} \{\hat{Y} = 1 \mid Y = 1, A = b\}$ \\

         &  & $\mathbb{P} \{\hat{Y} = 1 \mid Y = 0, A = a\} = \mathbb{P} \{\hat{Y} = 1 \mid Y = 0, A = b\}$ \\

           &  &  \\
       
       Sufficiency   & 
       $Y \perp A \mid \hat{Y}$ & 
       $\mathbb{P}  \{ Y = 1 \mid \hat{Y} = 1, A = a\} = \mathbb{P} \{Y = 1 \mid \hat{Y} = 1, A = b\}$ \\

         &  & $\mathbb{P} \{Y = 1 \mid \hat{Y} = 0, A = a\} = \mathbb{P} \{Y = 1 \mid \hat{Y} = 0, A = b\}$ \\ 

         &  &  \\
       
       \hline
    \end{tabular}
    \caption{Model Bias Criterion}
    \label{tab:bias_criterion}
\end{table}

Many statistical criteria have been proposed to determine whether a model is biased towards certain protected features.  We narrow our focus to the following three defined by Barocas, et al. \cite{BAROCAS2023}, and detailed in Table \ref{tab:bias_criterion}.

\begin{itemize}
    \item Independence
    \item Separation
    \item Sufficiency
\end{itemize}

Where random variable $A$ is a protected feature, $\hat{Y}$ is the binary classifier, and $Y$ is the target variable.  We also represent \textit{positive} as a {1} and \textit{negative} as a {0}. We choose to use the \textit{Independence} criterion, also known as \textit{disparate impact}, as it captures the notion of equal selection. Considering $\hat{Y}=1$ as \textit{selection}, the condition requires all groups to be accepted equally.  Allowing $a$ and $b$ to be swapped, the equation is often relaxed to meet the following ratio where $\epsilon$ is often taken to be $0.2$ \cite{FELDMAN2015}.

\begin{equation}
\label{equ:disparate_impact_theory}
    \textit{Disparate Impact} = \frac{\textit{selection rate for group $a$}}{\textit{selection rate for group $b$}} = \frac{\mathbb{P} \{\hat{Y} = 1 \mid A = a\}}{\mathbb{P} \{\hat{Y} = 1 \mid A = b\}}
    \geq 1 - \epsilon
\end{equation}

The \textit{disparate impact} can be used to evaluate model bias and considers how well the model performs for one group versus another group.  It is computed as a ratio, as shown in Equation \ref{equ:disparate_impact_theory}. We consider \textit{positive} to indicate the lesion is \textit{malignant}.  We define it in this manner because when the classifier fails to predict malignant it represents a lost opportunity (e.g. further tests, examinations, treatments).  Given this, the disparate impact can more intuitively be defined for our problem as shown in Equation \ref{equ:disparate_impact_tone}.


\begin{equation}
\textit{Tone Disparate Impact} = \frac{\textit{classifier predicts cancer for dark tone}}{\textit{classifier predicts cancer for light tone}}
\label{equ:disparate_impact_tone}
\end{equation}

We use the the \textit{Tone Disparate Impact} to evaluate our model for bias.


\subsection{Model Bias Solutions}
\label{sec:bias_solutions}

There are numerous proposed solutions to addressing model bias including regulatory and algorithmic.  Here we only consider algorithmic solutions to meet a criteria.  The algorithmic solutions generally fall into one of three techniques.

\begin{itemize}
    \item Preprocessing of the data
    \item Modify the model's weights during training
    \item Postprocessing (reweighing) the model predictions
\end{itemize}

Each has known strengths and weaknesses and will be explored in future work.  We provide here to facilitate our later discussion.



\subsection{Model Architecture}

Many models have been proposed for image classification, notably those based using Convolutional Neural Networks such Residual and Inception architectures \cite{HE2016}.  The canonical CNN model has a series of convolutional blocks, then a flatten layer, followed by series of linear blocks, and finally the prediction / output layer \cite{LECUN1998}.  To allow more inspection of the model, we define a custom CNN architecture for training and classifying images as either \textit{malignant} or \textit{benign}.  We note that we do not expect this custom model to outperform current techniques but it should outperform a majority class classifier.

\begin{figure}[h!]
\begin{center}
\includegraphics[width=16cm]{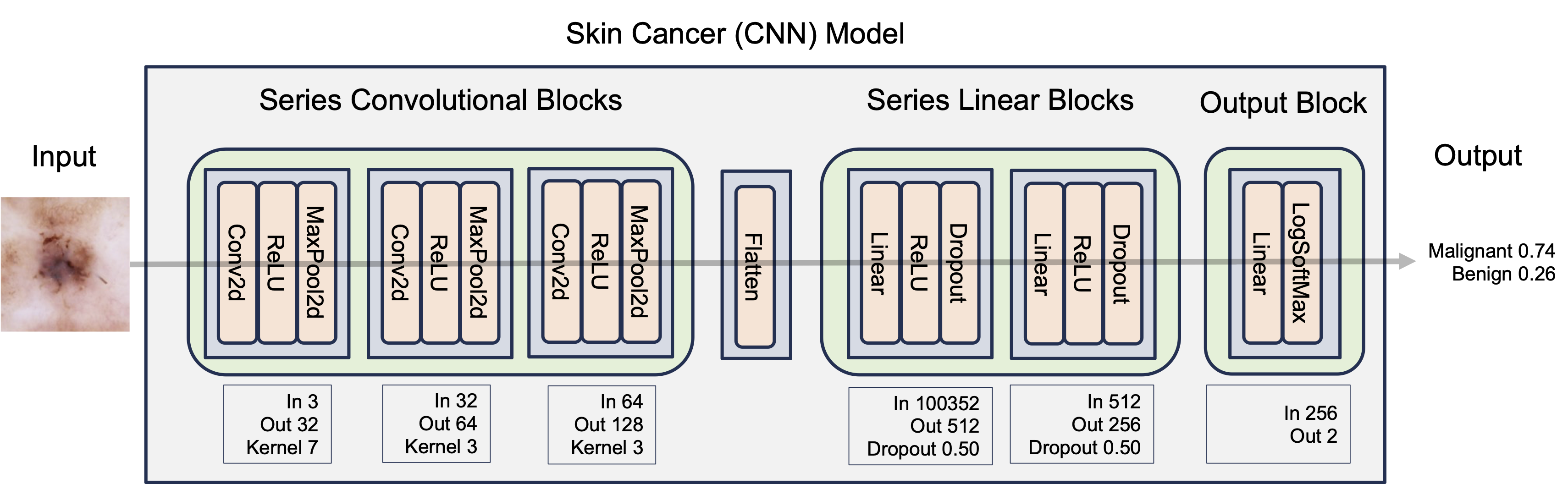}
\end{center}
\caption{Skin Cancer Image Classifier Architecture / Model}
\label{fig:overview_cnn}
\end{figure}

The initial layer takes the raw image of $\textit{width} \times \textit{height} \times \textit{features}$.  The features are also known as channels.  The input is expected to be $244 \times 244$ with 3 \textit{features}.  As the image is transformed through each block, the number of features increases while the image width and height reduce. Once the features are learned, they are flattened into one vector and input to a series of regular dense neural layers that can combine the low-level features into fewer, high-level features.  The final layer combines the high-level features using a soft max function to produce a probability distribution of the two classes.  This is shown in Figure \ref{fig:overview_cnn}.  

The convolutional blocks consist of a sequential convolutional layer (Conv2d), rectified linear unit activation (ReLU), and max pooling layer (MaxPool2d).  We chose the ReLU activation function as it is typical to use for image classification architectures \cite{IMAGENET2017}.  Each time the image is passed through the max pooling layer, the image width and height is reduced by two.  The linear blocks consists of a sequential dense layer (Linear), ReLU activation, and dropout layer.  The dropout layer has been shown to prevent overfitting and required some percentage of disabling the path/connection (simulating an ensemble of neural networks \cite{DROPOUT2014}.

\subsubsection{Hyper-parameter Tuning}

Deep learning architectures have a number of hyper-parameters that can have a significant effect on the model's prediction and computational training performance.  These hyper-parameters are often selected manually using the modeller's intuition or, more recently, using automated search methods. The presented architecture has the following hyper-parameters.  The ranges explored are given in the braces.

\begin{itemize}
    \item Number of convolutional blocks [2,6]
    \item Number of units for each convolutional layer [16,256]
    \item Number of linear blocks [2,6]
    \item Number of units for each linear layer [16,256]
    \item Percentage for each dropout layer [0.2,0.5]
    \item Learning rate step size [0.1, 0.00001]
    \item Optimiser \{Adam, SGD, RMSProp\}
\end{itemize}

Through a combination of automated (using the Optuna framework \cite{OPTUNA}) and manual hyper-parameter tuning we found three convolutional blocks and two linear blocks provided the best results.  This also informed our selection of blocks, units, and dropout percentage.  The tuning process also selected the Adam optimiser with a learning rate of 0.00001.

\subsection{Experimental Setup}

\begin{figure}
    \centering
    \includegraphics[width=\textwidth]{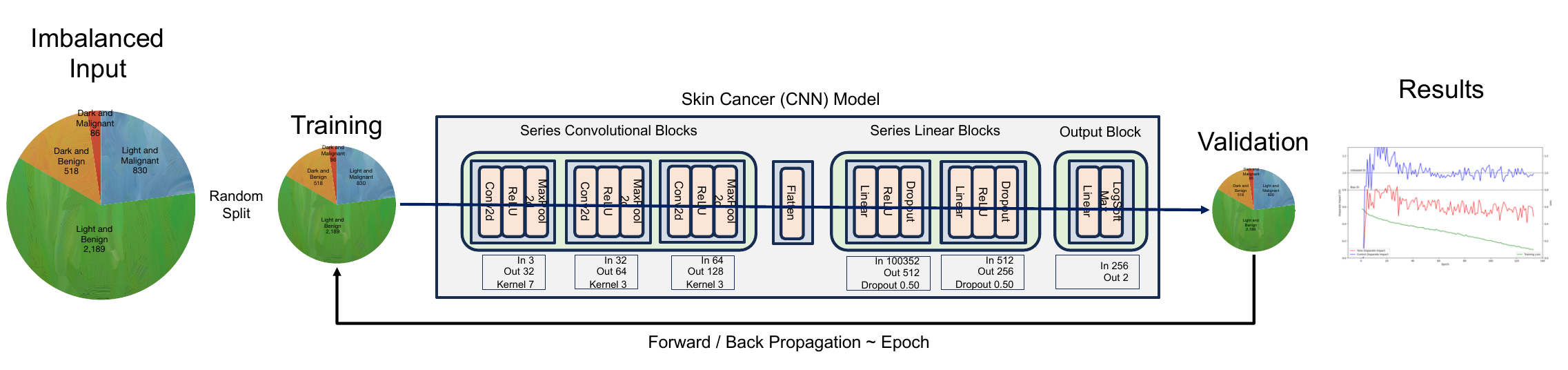}
    \caption{Imbalanced Experiment Setup}
    \label{fig:imbalanced_experiment}
\end{figure}

Figures \ref{fig:imbalanced_experiment} and \ref{fig:imbalanced_experiment} depict the  experimental setup.  The given input dataset is first randomly split into a training and validation set  (2/3 and 1/3 respectively).  For each epoch the training set is used in the forward pass through the model to produce predictions.  The predictions are compared against the validation set and results computed, to include the training loss, accuracy. and selection rates.  The backward pass then updates the model's weights.  This process is repeated each epoch with the results saved to be plotted and compared.

\begin{figure}
    \centering
    \includegraphics[width=\textwidth]{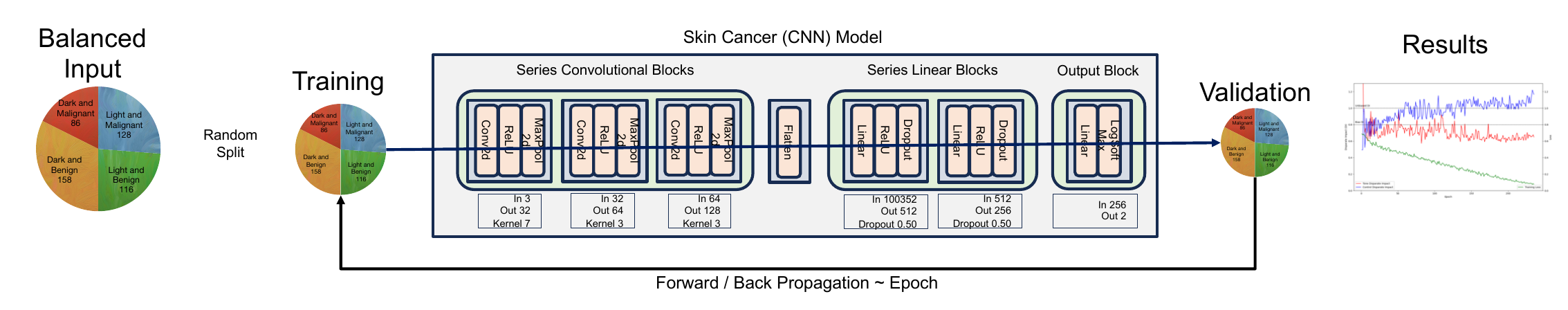}
    \caption{Balanced Experiment Setup}
    \label{fig:balanced_experiment}
\end{figure}

The models were implemented using the PyTorch deep learning framework  \cite{PYTORCH}. The experiments were conducted on both an Apple M2 Max system on a chip (12 CPU cores, 38 GPU cores, 16 neural engine cores, 64 GB RAM) and the Isambard-AI supercomputer (280 OpenMP cores, 916 GB RAM) \cite{ISAM2024}.  The Metal Performance Shaders (MPS) backend was used for the M2 Max.  For Isambard-AI the  OpenMP backend was used.


\section{Results}
\label{sec:results}
We first present the results for a model trained using the imbalanced dataset followed by the balanced dataset results. For each epoch a disparate impact is produced.  After presenting the disparate impact for all epochs, we examine the final epoch's confusion matrix and computation in more detail.

\subsection{Imbalanced Dataset Results}

To provide some indication that the disparate impact is not just spurious, we add a control feature to compare against.  The control feature has random values and, therefore, not correlated with the diagnosis (or skin tone).  We would expect the control to have a disparate impact between 0.8 and 1.2.  Anything outside (above or below) these thresholds would indicate bias.

\begin{figure}
\begin{center}
\includegraphics[width=\textwidth]{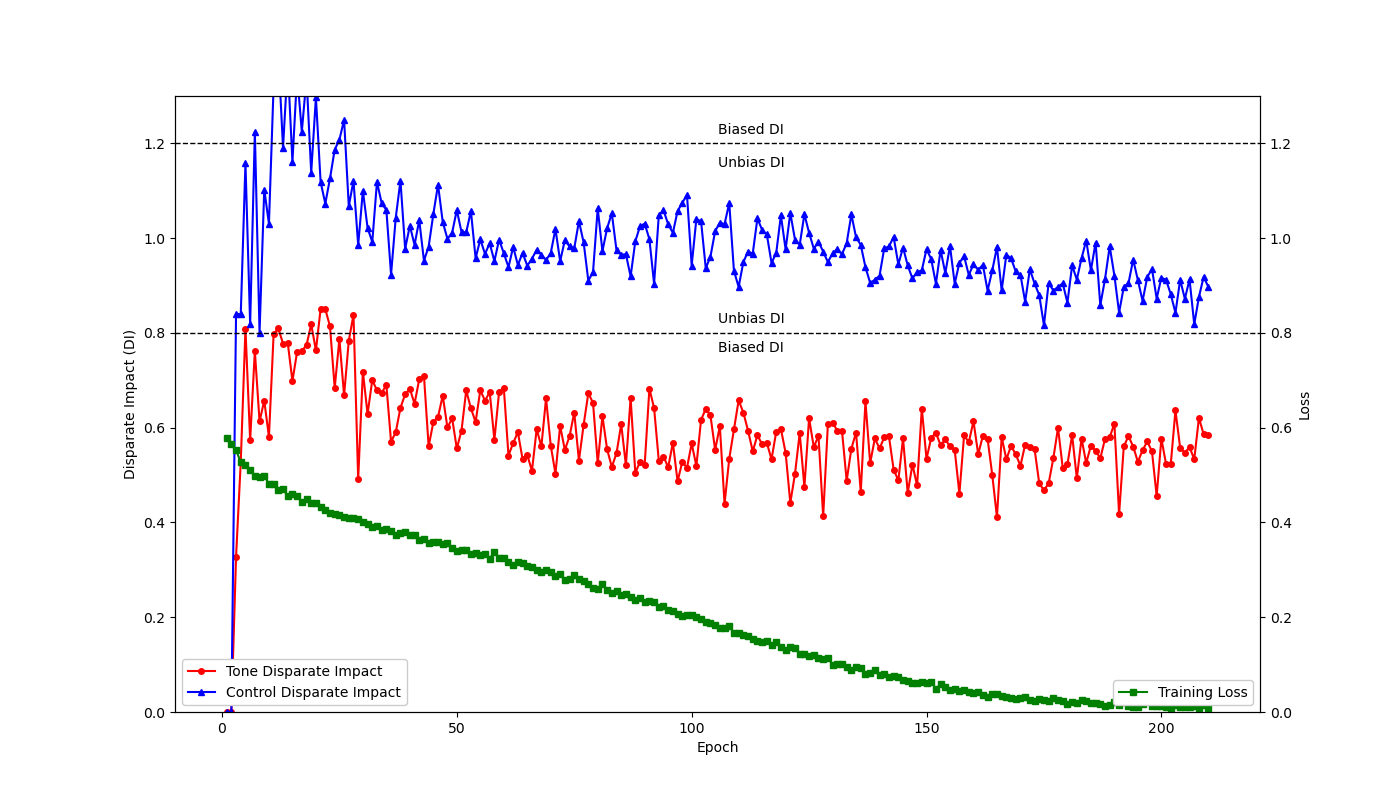}
\end{center}
\caption{Imbalanced Dataset Training and DI Curves}
\label{fig:imbalanced_training}
\end{figure}

We trained the model for approximately 200 epochs.   We compute the disparate impact for the tone and control features.  We also compute the training loss to provide an indication the model is learning. The tone and control disparate impact are shown from 0.0 to 1.3. The training loss is also (conveniently) shown in the same range. Figure \ref{fig:imbalanced_training} shows the results computed for each epoch.  We see significant variation in the early epochs when the model is not well trained.  After about 50 epochs the control remains within the expected, unbiased range.  We can clearly see that the tone disparate impact is consistently well below the threshold. Between 150 to 200 epochs, it appears the model has stopped learning (the training loss flattens).

To provide more context, we examine the final epoch's results. Table \ref{tab:imbalanced_cf_dark} shows the confusion matrix for the imbalanced model's predictions for dark tone images.  We can see that it predicts 23 positive images from the 189 in the validation set.

\begin{table}[!ht]
    \centering
    \begin{tabular}{l|l|c|c|c}
    \multicolumn{2}{c}{}&\multicolumn{2}{c}{True}&\\
    \cline{3-4}
    \multicolumn{2}{c|}{}&Positive&Negative&\multicolumn{1}{c}{Total}\\
    \cline{2-4}
    \multirow{2}{*}{Model}& Positive & $15$ & $15$ & $30$\\
    \cline{2-4}
    & Negative & $12$ & $147$ & $159$\\
    \cline{2-4}
    \multicolumn{1}{c}{} & \multicolumn{1}{c}{Total} & \multicolumn{1}{c}{$27$} & \multicolumn{    1}{c}{$162$} & \multicolumn{1}{c}{$189$}\\
    \end{tabular}
    \caption{Imbalanced Tone and Diagnosis Dataset Confusion Matrix (Epoch 200): Dark Tone Images}
    \label{tab:imbalanced_cf_dark}
\end{table}

Table \ref{tab:imbalanced_cf_light} shows the confusion matrix for the imbalanced model's predictions for light tone images.  The model predicts 187 positive images from the 898 in the validation set.

\begin{table}[!ht]
    \centering
    \begin{tabular}{l|l|c|c|c}
    \multicolumn{2}{c}{}&\multicolumn{2}{c}{True}&\\
    \cline{3-4}
    \multicolumn{2}{c|}{}&Positive&Negative&\multicolumn{1}{c}{Total}\\
    \cline{2-4}
    \multirow{2}{*}{Model}& Positive & $157$ & $90$ & $247$\\
    \cline{2-4}
    & Negative & $83$ & $568$ & $651$\\
    \cline{2-4}
    \multicolumn{1}{c}{} & \multicolumn{1}{c}{Total} & \multicolumn{1}{c}{$240$} & \multicolumn{    1}{c}{$658$} & \multicolumn{1}{c}{$898$}\\
    \end{tabular}
    \caption{Imbalanced Tone and Diagnosis Dataset Confusion Matrix (Epoch 200): Light Tone Images}
    \label{tab:imbalanced_cf_light}
\end{table}

Using the confusion matrix results, Equation \ref{equ:disparate_impact_tone_im} shows that the disparate impact is 0.584. This is the Tone Disparate Impact value shown near the last epoch of Figure \ref{fig:imbalanced_training}.

\begin{equation}
\textit{Tone Disparate Impact}_{Imbalanced} = \frac{\textit{classifier predicts cancer for dark tone}}{\textit{classifier predicts cancer for light tone}} = \frac{30 / 189}{247 / 898} = 0.577
\label{equ:disparate_impact_tone_im}
\end{equation}

\subsection{Balanced Dataset Results}

We trained the balanced model for approximately 350 epochs. This was trained for more epochs than the imbalanced case as it took more epochs before the training loss flattened (around epoch 300).  Figure \ref{fig:imbalanced_training} shows the results computed for each epoch.  Similar to the imbalanced results, we see significant variation of the disparate impact in the early epochs.  After 50 epochs the control remains within the expected range but the tone remains below the 0.80 threshold, though not as significant as with the imbalanced disparate impact.

\begin{figure}[!ht]
\begin{center}
\includegraphics[width=\textwidth]{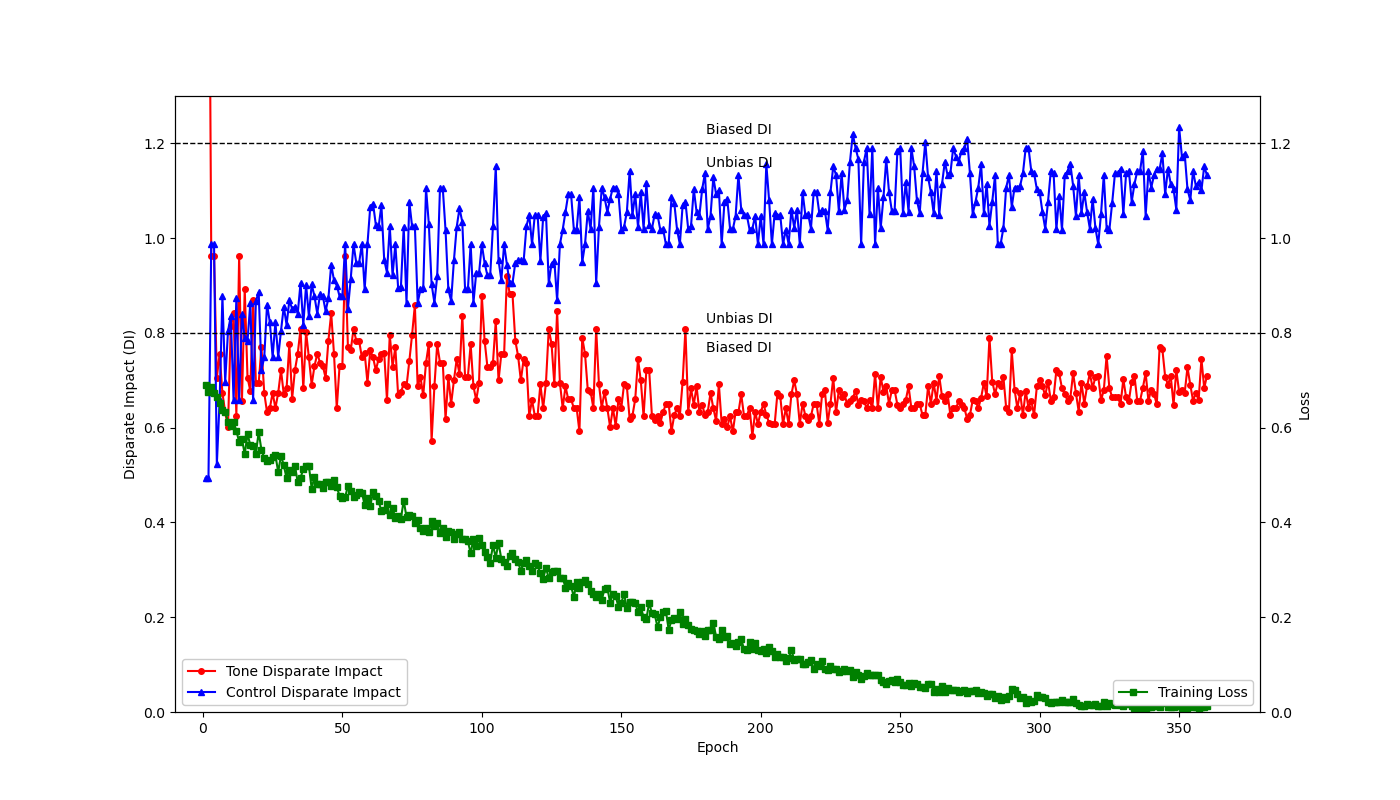}
\end{center}
\caption{Balanced Dataset Training and DI Curves}
\label{fig:balanced_training}
\end{figure}

To provide additional context, we examine the final epoch's results. Table \ref{tab:balanced_cf_dark} shows the confusion matrix for the balanced model's predictions for dark tone images.  We can see that it predicts 31 positive images from the 79 in the validation set.

\begin{table}[!ht]
    \centering
    \begin{tabular}{l|l|c|c|c}
    \multicolumn{2}{c}{}&\multicolumn{2}{c}{True}&\\
    \cline{3-4}
    \multicolumn{2}{c|}{}&Positive&Negative&\multicolumn{1}{c}{Total}\\
    \cline{2-4}
    \multirow{2}{*}{Model}& Positive & $14$ & $13$ & $27$\\
    \cline{2-4}
    & Negative & $10$ & $42$ & $52$\\
    \cline{2-4}
    \multicolumn{1}{c}{} & \multicolumn{1}{c}{Total} & \multicolumn{1}{c}{$24$} & \multicolumn{    1}{c}{$55$} & \multicolumn{1}{c}{$79$}\\
    \end{tabular}
    \caption{Balanced Tone and Diagnosis Dataset Confusion Matrix (Epoch 325): Dark Tone Images}
    \label{tab:balanced_cf_dark}
\end{table}

Table \ref{tab:balanced_cf_light} shows the confusion matrix for the balanced model's predictions for light tone images.  The model predicts 42 positive images from the 76 in the validation set.

\begin{table}[!ht]
    \centering
    \begin{tabular}{l|l|c|c|c}
    \multicolumn{2}{c}{}&\multicolumn{2}{c}{True}&\\
    \cline{3-4}
    \multicolumn{2}{c|}{}&Positive&Negative&\multicolumn{1}{c}{Total}\\
    \cline{2-4}
    \multirow{2}{*}{Model}& Positive & $32$ & $10$ & $42$\\
    \cline{2-4}
    & Negative & $12$ & $22$ & $34$\\
    \cline{2-4}
    \multicolumn{1}{c}{} & \multicolumn{1}{c}{Total} & \multicolumn{1}{c}{$44$} & \multicolumn{    1}{c}{$32$} & \multicolumn{1}{c}{$76$}\\
    \end{tabular}
    \caption{Balanced Tone and Diagnosis Dataset Confusion Matrix (Epoch 325): Light Tone Images}
    \label{tab:balanced_cf_light}
\end{table}

Using the confusion matrix results, Equation \ref{equ:disparate_impact_tone_ba} shows that the disparate impact is 0.684. This is the Tone Disparate Impact value shown near the last epoch in Figure \ref{fig:balanced_training}.

\begin{equation}
\textit{Tone Disparate Impact}_{Balanced} = \frac{\textit{classifier predicts cancer for dark tone}}{\textit{classifier predicts cancer for light tone}} = \frac{27 / 79}{38 / 76} = 0.684
\label{equ:disparate_impact_tone_ba}
\end{equation}


\subsection{Model Accuracy Results}
\label{sec:model_accuracy}

Table \ref{tab:accuracy_comparison} shows the accuracy for the balanced and imbalanced models (technically the models trained using the imbalanced and balanced datasets) computed using Equation \ref{equ:accuracy}.  Note that the accuracy is derived from the last epoch after training loss has stopped decreasing.  We compare against the majority class classifier derived from Figures \ref{fig:imbalanced} and \ref{fig:balanced}.

\begin{table}
\centering
\caption{Model Accuracy Comparison (Final Epoch)}
\label{tab:accuracy_comparison}
\begin{tabular}[t]{lcc}
\toprule
Model &\# Accuracy & Majority Classifier Accuracy\\
\midrule
Imbalanced Model & 0.83 & 0.74 \\
Balanced Model & 0.72 & 0.55 \\
\bottomrule
\end{tabular}
\end{table}

\section{Discussion}
\label{sec:discussion}
\subsection{Disparate Impact for Imbalanced and Balanced Models}

The balanced model resulted in a disparate impact of 0.71.  The imbalanced model resulted in a disparate impact of approximately 0.58.  Regardless of the relative number of malignant versus benign diagnosed images or relative number of light versus dark tone images, the disparate impact indicates the model is better at selecting light tone images versus dark tone images.  The disparate impact for both imbalanced and balanced datasets is below the 0.80 threshold indicating modal bias.  

The results suggest that the balanced model results may have less bias than the imbalanced model.  However, there is clearly randomness in the disparate impact as the model's are trained each epoch. We did repeat these experiments and confirmed similar results where the disparate impact of the balanced model was less than the imbalanced model. However, more experiments and tests are necessary to confirm this hypothesis and will be reported in future work. 

\subsection{Dataset Size, Class Imbalance, and Model Accuracy}

We chose to address the class imbalance by under sampling \textit{benign} diagnosis and then under sampling \textit{light} tone images.  Of the 3,623 images, this left roughly 500 images to train and test the models. Deep learning models typically require large amounts of data to achieve high accuracy.  Our model's accuracy was not significantly larger than the majority class classifier, likely due to the fact this there was minimal training data.

With the 3,623 images, the model accuracy was reasonable.  We are careful to note that many high accuracy models include other features, such as exposure to the sun, location of the lesion, etc. \cite{HAM2018}.

One solution to increase the number of images is to develop a tone classifier to label the skin tone of the remaining 78,000 images.  Another, possibly more controversial, approach would be to use generative AI approaches to create more \textit{malignant} and \textit{dark} images.

\subsection{Model Bias Evaluation}

As discussed in Section \ref{sec:bias_evaluation}, there are several criterion to evaluate model bias.  It can be shown that satisfying one of the three criteria presented by Barocas, et al. \cite{BAROCAS2023}, will violate one or both the other criterion.  We chose to evaluate the model using \textit{Independence} (via the \textit{disparate impact} measure).  If we used one of the solutions mentioned in Section \ref{sec:bias_solutions} to meet this criteria, we would necessarily violate the \textit{Separation} criteria.  Unfortunately, this means we cannot simply declare our model "free from bias" given we satisfy one of these criteria.  Nevertheless, not meeting any criteria or being completely unaware of a model's bias is less desirable.

Balancing the dataset for tone and diagnosis is considered a data preprocessing solution to address the bias.  Though this did not work, we believe further data preprocessing techniques are important to consider.  This motivates exploration of other techniques (outlined in Section \ref{sec:bias_solutions}) to mitigate the model bias.

\subsection{Skin Tone Definition}

We somewhat arbitrary decided to delineate light and dark tone using the Fitzpatrick skin types.  We note there are no V or VI skin types in ISIC archive.  There are other definitions of skin tone and future work will explore defining light and dark using these definitions \cite{MINDGAP2020}.

\subsection{Model Accuracy Comparison}

The model accuracy results were compared against the majority class classifier to ensure the model is learning how to differentiate the diagnosis versus just  guess randomly or guessing the majority class (i.e. benign).  Though both models performed better than the majority classifier, the model trained on the balanced dataset performed relatively better.  However, these accuracy results are well below state-of-the-art \cite{BRINKER2018} for more sophisticated deep neural networks trained on much larger datasets.  We believe this is due to the relatively small datasets used to train our models. This work was limited due to the lack of skin type annotations.  Future work will consider these more sophisticated models with larger datasets derived using a tone classifier.

\section{Acknowledgements}
\label{sec:ack}
This work was initially support by a Jean Golding Institute, University of Bristol, Seed Corn Grant "Pilot Study to determine Tone Bias in Open-Source Skin Cancer Datasets" awarded 8 November 2023.  Subsequent computational support was provided by Isambard-AI, Bristol Centre for Supercomputing \cite{ISAM2024}.  The authors would like to thank the Jean Golding Institute's Dr. Huw Day and Dr. Will Chapman for their support throughout the project.  We would also like to thank Christopher Woods for support using the Isambard supercomputer.

\bibliographystyle{unsrtnat}
\bibliography{references}

\begin{thebibliography}{24}
\providecommand{\natexlab}[1]{#1}
\providecommand{\url}[1]{\texttt{#1}}
\expandafter\ifx\csname urlstyle\endcsname\relax
  \providecommand{\doi}[1]{doi: #1}\else
  \providecommand{\doi}{doi: \begingroup \urlstyle{rm}\Url}\fi

\bibitem[Cassidy et~al.(2022)Cassidy, Kendrick, Brodzicki, Jaworek-Korjakowska, and Yap]{CASSIDY2022}
Bill Cassidy, Connah Kendrick, Andrzej Brodzicki, Joanna Jaworek-Korjakowska, and Moi~Hoon Yap.
\newblock Analysis of the isic image datasets: Usage, benchmarks and recommendations.
\newblock \emph{Medical Image Analysis}, 75:\penalty0 102305, 2022.
\newblock ISSN 1361-8415.
\newblock \doi{https://doi.org/10.1016/j.media.2021.102305}.
\newblock URL \url{https://www.sciencedirect.com/science/article/pii/S1361841521003509}.

\bibitem[Jain et~al.(2021)Jain, Way, Gupta, Gao, de~Oliveira~Marinho, Hartford, Sayres, Kanada, Eng, Nagpal, DeSalvo, Corrado, Peng, Webster, Dunn, Coz, Huang, Liu, Bui, and Liu]{JAIN2021}
Ayush Jain, David Way, Vishakha Gupta, Yi~Gao, Guilherme de~Oliveira~Marinho, Jay Hartford, Rory Sayres, Kimberly Kanada, Clara Eng, Kunal Nagpal, Karen~B. DeSalvo, Greg~S. Corrado, Lily Peng, Dale~R. Webster, R.~Carter Dunn, David Coz, Susan~J. Huang, Yun Liu, Peggy Bui, and Yuan Liu.
\newblock {Development and Assessment of an Artificial Intelligence–Based Tool for Skin Condition Diagnosis by Primary Care Physicians and Nurse Practitioners in Teledermatology Practices}.
\newblock \emph{JAMA Network Open}, 4\penalty0 (4):\penalty0 e217249--e217249, 04 2021.
\newblock ISSN 2574-3805.
\newblock \doi{10.1001/jamanetworkopen.2021.7249}.
\newblock URL \url{https://doi.org/10.1001/jamanetworkopen.2021.7249}.

\bibitem[Liu et~al.(2020)Liu, Jain, Eng, Way, Lee, Bui, Kanada, de~Oliveira~Marinho, Gallegos, Gabriele, Gupta, Singh, Natarajan, Hofmann-Wellenhof, Corrado, Peng, Webster, Ai, Huang, Liu, Dunn, and Coz]{LIU2020}
Yuan Liu, Ayush Jain, Clara Eng, David~H. Way, Kang Lee, Peggy Bui, Kimberly Kanada, Guilherme de~Oliveira~Marinho, Jessica Gallegos, Sara Gabriele, Vishakha Gupta, Nalini Singh, Vivek Natarajan, Rainer Hofmann-Wellenhof, Greg~S. Corrado, Lily~H. Peng, Dale~R. Webster, Dennis Ai, Susan~J. Huang, Yun Liu, R.~Carter Dunn, and David Coz.
\newblock A deep learning system for differential diagnosis of skin diseases.
\newblock \emph{Nature Medicine}, 26\penalty0 (6):\penalty0 900--908, Jun 2020.
\newblock ISSN 1546-170X.
\newblock \doi{10.1038/s41591-020-0842-3}.
\newblock URL \url{https://doi.org/10.1038/s41591-020-0842-3}.

\bibitem[Wen et~al.(2022)Wen, Khan, Ji~Xu, Ibrahim, Smith, Caballero, Zepeda, de~Blas~Perez, Denniston, Liu, and Matin]{LANCET2022}
David Wen, Saad~M. Khan, Antonio Ji~Xu, Hussein Ibrahim, Luke Smith, Jose Caballero, Luis Zepeda, Carlos de~Blas~Perez, Alastair~K. Denniston, Xiaoxuan Liu, and Rubeta~N. Matin.
\newblock Characteristics of publicly available skin cancer image datasets: a systematic review.
\newblock \emph{The Lancet Digital Health}, 4\penalty0 (1):\penalty0 e64--e74, Jan 2022.
\newblock ISSN 2589-7500.
\newblock \doi{10.1016/S2589-7500(21)00252-1}.
\newblock URL \url{https://doi.org/10.1016/S2589-7500(21)00252-1}.

\bibitem[Green et~al.(2024)Green, Murphy, and Robinson]{AIBIAS2024}
B.~Lee Green, Anastasia Murphy, and Edmondo Robinson.
\newblock Accelerating health disparities research with artificial intelligence.
\newblock \emph{Frontiers in Digital Health}, 6, 2024.
\newblock ISSN 2673-253X.
\newblock \doi{10.3389/fdgth.2024.1330160}.
\newblock URL \url{https://www.frontiersin.org/journals/digital-health/articles/10.3389/fdgth.2024.1330160}.

\bibitem[Barocas et~al.(2023)Barocas, Hardt, and Narayanan]{BAROCAS2023}
Solon Barocas, Moritz Hardt, and Arvind Narayanan.
\newblock \emph{Fairness and Machine Learning: Limitations and Opportunities}.
\newblock MIT Press, 2023.

\bibitem[Bonham et~al.(2016)Bonham, Callier, and {Royal, Charmaine D}]{BONHAM2016}
Vence~L Bonham, Shawneequa~L Callier, and {Royal, Charmaine D}.
\newblock Will precision medicine move us beyond race?
\newblock \emph{N Engl J Med}, 374\penalty0 (21):\penalty0 2003--2005, May 2016.

\bibitem[Mukwende et~al.(2020)Mukwende, Tamony, and Turner]{MINDGAP2020}
Malone Mukwende, Peter Tamony, and Margot Turner.
\newblock \emph{{Mind the Gap: A handbook of clinical signs in Black and Brown skin}}.
\newblock St Georges University of London, 8 2020.
\newblock \doi{10.24376/rd.sgul.12769988.v1}.

\bibitem[Barata et~al.(2023)Barata, Rotemberg, Codella, Tschandl, Rinner, Akay, Apalla, Argenziano, Halpern, Lallas, Longo, Malvehy, Puig, Rosendahl, Soyer, Zalaudek, and Kittler]{BARATA2023}
Catarina Barata, Veronica Rotemberg, Noel C~F Codella, Philipp Tschandl, Christoph Rinner, Bengu~Nisa Akay, Zoe Apalla, Giuseppe Argenziano, Allan Halpern, Aimilios Lallas, Caterina Longo, Josep Malvehy, Susana Puig, Cliff Rosendahl, H~Peter Soyer, Iris Zalaudek, and Harald Kittler.
\newblock A reinforcement learning model for {AI-based} decision support in skin cancer.
\newblock \emph{Nat Med}, 29\penalty0 (8):\penalty0 1941--1946, July 2023.

\bibitem[Jain(2024)]{JAIN2024}
Tanish Jain.
\newblock Evaluating machine learning-based skin cancer diagnosis, 2024.
\newblock URL \url{https://arxiv.org/abs/2409.03794}.

\bibitem[Gupta and Sharma(2019)]{GUPTA2019}
Vishal Gupta and Vinod~Kumar Sharma.
\newblock Skin typing: Fitzpatrick grading and others.
\newblock \emph{Clinics in Dermatology}, 37\penalty0 (5):\penalty0 430--436, 2019.
\newblock ISSN 0738-081X.
\newblock \doi{https://doi.org/10.1016/j.clindermatol.2019.07.010}.
\newblock URL \url{https://www.sciencedirect.com/science/article/pii/S0738081X1930121X}.
\newblock The Color of Skin.

\bibitem[Codella et~al.(2018)Codella, Gutman, Celebi, Helba, Marchetti, Dusza, Kalloo, Liopyris, Mishra, Kittler, and Halpern]{ISIC_Archive}
Noel C.~F. Codella, David Gutman, M.~Emre Celebi, Brian Helba, Michael~A. Marchetti, Stephen~W. Dusza, Aadi Kalloo, Konstantinos Liopyris, Nabin Mishra, Harald Kittler, and Allan Halpern.
\newblock Skin lesion analysis toward melanoma detection: A challenge at the 2017 international symposium on biomedical imaging (isbi), hosted by the international skin imaging collaboration (isic), 2018.
\newblock URL \url{https://arxiv.org/abs/1710.05006}.

\bibitem[Adhikari et~al.(2024)Adhikari, Martyshkin, Fedorov, Das, Antony, and Mirov]{NIRM2024}
Nirmala Adhikari, Dmitry Martyshkin, Vladimir Fedorov, Deblina Das, Veena Antony, and Sergey Mirov.
\newblock Laser-induced breakdown spectroscopy detection of heavy metal contamination in soil samples from north birmingham, alabama.
\newblock \emph{Applied Sciences}, 14\penalty0 (17), 2024.
\newblock ISSN 2076-3417.
\newblock \doi{10.3390/app14177868}.
\newblock URL \url{https://www.mdpi.com/2076-3417/14/17/7868}.

\bibitem[Chardon et~al.(1991)Chardon, Cretois, and Hourseau]{CHARDON1991}
A~Chardon, I~Cretois, and C~Hourseau.
\newblock Skin colour typology and suntanning pathways.
\newblock \emph{Int J Cosmet Sci}, 13\penalty0 (4):\penalty0 191--208, August 1991.

\bibitem[Feldman et~al.(2015)Feldman, Friedler, Moeller, Scheidegger, and Venkatasubramanian]{FELDMAN2015}
Michael Feldman, Sorelle~A. Friedler, John Moeller, Carlos Scheidegger, and Suresh Venkatasubramanian.
\newblock Certifying and removing disparate impact.
\newblock In \emph{Proceedings of the 21th ACM SIGKDD International Conference on Knowledge Discovery and Data Mining}, KDD '15, page 259–268, New York, NY, USA, 2015. Association for Computing Machinery.
\newblock ISBN 9781450336642.
\newblock \doi{10.1145/2783258.2783311}.
\newblock URL \url{https://doi.org/10.1145/2783258.2783311}.

\bibitem[He et~al.(2016)He, Zhang, Ren, and Sun]{HE2016}
Kaiming He, Xiangyu Zhang, Shaoqing Ren, and Jian Sun.
\newblock Deep residual learning for image recognition.
\newblock In \emph{2016 IEEE Conference on Computer Vision and Pattern Recognition (CVPR)}, pages 770--778, 2016.
\newblock \doi{10.1109/CVPR.2016.90}.

\bibitem[Lecun et~al.(1998)Lecun, Bottou, Bengio, and Haffner]{LECUN1998}
Y.~Lecun, L.~Bottou, Y.~Bengio, and P.~Haffner.
\newblock Gradient-based learning applied to document recognition.
\newblock \emph{Proceedings of the IEEE}, 86\penalty0 (11):\penalty0 2278--2324, 1998.
\newblock \doi{10.1109/5.726791}.

\bibitem[Krizhevsky et~al.(2017)Krizhevsky, Sutskever, and Hinton]{IMAGENET2017}
Alex Krizhevsky, Ilya Sutskever, and Geoffrey~E. Hinton.
\newblock Imagenet classification with deep convolutional neural networks.
\newblock \emph{Commun. ACM}, 60\penalty0 (6):\penalty0 84–90, May 2017.
\newblock ISSN 0001-0782.
\newblock \doi{10.1145/3065386}.
\newblock URL \url{https://doi.org/10.1145/3065386}.

\bibitem[Srivastava et~al.(2014)Srivastava, Hinton, Krizhevsky, Sutskever, and Salakhutdinov]{DROPOUT2014}
Nitish Srivastava, Geoffrey Hinton, Alex Krizhevsky, Ilya Sutskever, and Ruslan Salakhutdinov.
\newblock Dropout: A simple way to prevent neural networks from overfitting.
\newblock \emph{Journal of Machine Learning Research}, 15\penalty0 (56):\penalty0 1929--1958, 2014.
\newblock URL \url{http://jmlr.org/papers/v15/srivastava14a.html}.

\bibitem[Akiba et~al.(2019)Akiba, Sano, Yanase, Ohta, and Koyama]{OPTUNA}
Takuya Akiba, Shotaro Sano, Toshihiko Yanase, Takeru Ohta, and Masanori Koyama.
\newblock Optuna: A next-generation hyperparameter optimization framework.
\newblock In \emph{Proceedings of the 25th ACM SIGKDD International Conference on Knowledge Discovery \& Data Mining}, KDD '19, page 2623–2631, New York, NY, USA, 2019. Association for Computing Machinery.
\newblock ISBN 9781450362016.
\newblock \doi{10.1145/3292500.3330701}.
\newblock URL \url{https://doi.org/10.1145/3292500.3330701}.

\bibitem[Paszke et~al.(2019)Paszke, Gross, Massa, Lerer, Bradbury, Chanan, Killeen, Lin, Gimelshein, Antiga, Desmaison, Köpf, Yang, DeVito, Raison, Tejani, Chilamkurthy, Steiner, Fang, Bai, and Chintala]{PYTORCH}
Adam Paszke, Sam Gross, Francisco Massa, Adam Lerer, James Bradbury, Gregory Chanan, Trevor Killeen, Zeming Lin, Natalia Gimelshein, Luca Antiga, Alban Desmaison, Andreas Köpf, Edward Yang, Zach DeVito, Martin Raison, Alykhan Tejani, Sasank Chilamkurthy, Benoit Steiner, Lu~Fang, Junjie Bai, and Soumith Chintala.
\newblock Pytorch: An imperative style, high-performance deep learning library, 2019.
\newblock URL \url{https://arxiv.org/abs/1912.01703}.

\bibitem[McIntosh-Smith et~al.(2024)McIntosh-Smith, Alam, and Woods]{ISAM2024}
Simon McIntosh-Smith, Sadaf~R Alam, and Christopher Woods.
\newblock {Isambard-AI: a leadership class supercomputer optimised specifically for Artificial Intelligence}, 2024.
\newblock URL \url{https://arxiv.org/abs/2410.11199}.

\bibitem[Tschandl et~al.(2018)Tschandl, Rosendahl, and Kittler]{HAM2018}
Philipp Tschandl, Cliff Rosendahl, and Harald Kittler.
\newblock The {HAM10000} dataset, a large collection of multi-source dermatoscopic images of common pigmented skin lesions.
\newblock \emph{Scientific Data}, 5\penalty0 (1):\penalty0 180161, August 2018.

\bibitem[Brinker et~al.(2018)Brinker, Hekler, Utikal, Grabe, Schadendorf, Klode, Berking, Steeb, Enk, and von Kalle]{BRINKER2018}
Titus~Josef Brinker, Achim Hekler, Jochen~Sven Utikal, Niels Grabe, Dirk Schadendorf, Joachim Klode, Carola Berking, Theresa Steeb, Alexander~H Enk, and Christof von Kalle.
\newblock Skin cancer classification using convolutional neural networks: Systematic review.
\newblock \emph{J Med Internet Res}, 20\penalty0 (10):\penalty0 e11936, October 2018.

\end{thebibliography}

\end{document}